\documentstyle[12pt]{article}
\begin{document}
\begin{titlepage}
\begin{flushright}
TU-553
\end{flushright}
\ \\
\ \\
\ \\
\ \\
\begin{center}
{\LARGE \bf
Asymptotic Isometry \\
and \\
 Two Dimensional Anti-de Sitter Gravity \\
}
\end{center}
\ \\
\ \\
\begin{center}
\Large{
M.Hotta
 }\\
{\it 
Department of Physics, Tohoku University,\\
Sendai 980-8578, Japan
}
\end{center}
\ \\
\ \\

\begin{abstract}
In low dimensional gravity on anti-de sitter space   
 there exists a  possibility that the asymptotic isometry 
  is raised to a Virasoro symmetry living at spatial infinity. 
We discuss in detail asymptotic isometry of the most general  
anti-de sitter dilaton gravity in two dimensions. 
 From  analysis of the boundary dynamics 
  it turns out  that the Virasoro symmetry 
is not preserved in all the models. 
 Moreover even anti-de sitter isomery cannot survive 
and asymptotic isometry consists of only time translation  
for all the models except Jackiw-Teitelboim (JT) model. 
 Meanwhile there exists 
a hybride nonsinglet representation of the isometry 
 in JT model.
\end{abstract}

\end{titlepage}

\section{
Introduction
}

\ \\

The conception of asymptotic isometry of anti-de sitter (AdS) space
 has recently attracted much attention from the 
 AdS/CFT correspondence viewpoint\cite{M} 
and gives a lot of information about
 the black hole entropy problem, strong coupling behavior
 of gauge theory and string theory\cite{S,W}. 
Here AdS/CFT correspondence means that  supergravity 
on AdS space is equivalent with a 
conformal field theory on the boundary.

Originally the asymptotic isometry of  three dimensional 
anti-de sitter space  
 was discussed in 1986 by Brown and Henneaux\cite{BH}. They showed
 that the asymptotic symmetry of Einsiten gravity on AdS space   
 is not just the $SO(2,2)$ AdS isometry but rather 
a two dimensional conformal (Virasoro) symmetry at spatial infinity
 which includes the isometry as a subgroup. 
They also identified the central charge
of the Virasoro algebra with
$$
c=\frac{3}{2G\sqrt{-\Lambda}}
$$
 where $G$ is the gravitational constant and $\Lambda$
 negative cosmological constant.

In two dimensions there exists a possibility that
 the  AdS asymptotic isometry is 
also raised to a conformal symmetry  
on the boundary just like in three dimensions.

In this paper we analyze asymptotic isometry of two dimensional
 gravity in detail.  
 We focus on AdS dilaton gravity models
 in which all the spacetimes of solutions possess a negative 
constant curvature. The AdS gravity models include  Liouville
 induced gravity model\cite{P} and Jackiw-Teitelboim (JT) model
\cite{JT} with negative cosmological constant.

We prove that nontrivial asymptotic isometry 
of general AdS dilaton gravity models except JT model 
 is only the time translation. If the conformal symmetry
 on the boundary  is respected, 
the canonical representation is forced to be singlet.

On the other hand,  the situation slightly changes in JT model.
 We show that the AdS symmetry can survive as asymptotic isometry
in a nonsinglet representation composed of different  
canonical representation sectors.

 Our result suggests that  
 even when the models are extented to supergravity and quantized
 the asymptotic states do not belong to any representaion
 of the boundary conformal group and 
that AdS/CFT correspondence does not work well in two dimensions. 
 This aspect is quite different from in three dimensions.

Here we want to comment on relations with other works already 
performed. 
 A canonical representation of JT model has been already 
discussed by Henneaux\cite{H}. However he concentrated on
 an AdS variant boundary condition and did not deal with 
  any combined representations to recover the AdS symmetry.

Quantization of JT model has been also 
investigated\cite{H,T}. However 
our nonsinglet representation has not yet been treated explictly . 
In this paper quantum realization of the AdS symmetry of JT model 
is beyond the scope.

\section{AdS Space in Two Dimensions}
\ \\

Let us first review in brief two dimensional AdS space.
This can be expressed as an embedded hypersurface in three
 dimensional flat spacetime with its metric signature 
$(-1,-1,1)$.
\begin{eqnarray}
&&
-T^2_1 -T^2_2 +X^2 =-\frac{1}{\mu^2},\label{100}
\\
&&
ds^2=-dT^2_1 -dT^2_2 +dX^2 .\label{33}
\end{eqnarray}
Here $\mu$ is a constant related with the curvature 
$R=-2\mu^2$.

Taking the following parametrization:
\begin{eqnarray}
&&
T_1 =\frac{1}{\mu}\frac{\cos(\mu t)}{\sin(\mu x)},
\\
&&
T_2 =\frac{1}{\mu}\frac{\sin(\mu t)}{\sin(\mu x)},
\\
&&
X=\frac{1}{\mu}\cot(\mu x),
\end{eqnarray}
 a metric in the full covered chart is obtained as
\begin{eqnarray}
ds^2 =\frac{1}{\sin^2 (\mu x)}(-dt^2 +dx^2), \label{1}
\end{eqnarray}
where the coordinate variables run between 
\begin{eqnarray}
&&
-\infty <t <\infty,
\\
&&
-\frac{\pi}{\mu} < x < 0.
\end{eqnarray}
The spatial infinity stays at $x=0$ and $x=-\frac{\pi}{\mu}$.
In the following discussion, we will discuss asymptotic AdS space
in which the space approaches to AdS space near $x\sim -0$.
 So omit another infinity and spacetimes 
with no asymptotic region on the left side can be taken account of.

It is a well known fact that 
the metric eqn(\ref{1}) has three killing vectors which 
 forms $SO(1,2)$ group and its infinitesimal 
transformation can be explicitly written as follows.
\begin{eqnarray}
&&
x^{'\mu}=x^\mu -\epsilon^\mu (x), \label{3}
\\
&&
\epsilon^t = \epsilon_o +\epsilon_1 \cos (\mu x)
 \sin(\mu t +\delta_\epsilon ) ,\label{34}
\\
&&
\epsilon^x =
\epsilon_1 \sin(\mu x) \cos(\mu t+\delta_\epsilon), \label{35}
\end{eqnarray}
where $\epsilon_0$, $\epsilon_1$ 
and $\delta_\epsilon$ are group element parameters.
These equations  are also rewritten 
in a compact form using the conformal
coordinates as 
\begin{eqnarray}
&&
x^\pm =t\pm x ,
\\
&&
\epsilon^\pm =\epsilon_0  +\epsilon_1 
\sin\left(\mu x^\pm +\delta_\epsilon  \right) \label{501} .
\end{eqnarray}
Generators of these transformations are defined as
\begin{eqnarray}
&&
L_0 =\frac{i}{\mu}\partial_x,
\\
&&
L_1= \frac{i}{\mu}e^{i\mu x} \partial_x ,
\\
&&
L_{-1} =\frac{i}{\mu}e^{-i\mu x} \partial_x ,
\end{eqnarray}
and form the SO(1,2) algebra: 
\begin{eqnarray}
&&
[L_0 ,\ L_{\pm 1} ]=\mp L_{\pm 1},
\\
&&
[L_1 ,\ L_{-1}]=2L_0.
\end{eqnarray}
This isometry is called AdS symmtery and has a root in 
 maximally symmetric nature of the AdS space. 
The algebra has a casimir operator which is defined 
to commute with all the generators, and  is expressed as
\begin{eqnarray}
M^2 =\mu^2 
\left[
L_0^2 -\frac{1}{2}
\left( 
L_1 L_{-1} +L_{-1} L_1
\right)
\right].
\end{eqnarray}

There is another coordinate system frequently used
 because of simplicity of the metric form. 
Substitution  of a parametrization form:
\begin{eqnarray}
&&
T_1 = -\frac{1}{2\mu}
\left( 
\frac{1}{\mu x} +\mu x
\right)
+\frac{t^2}{2x},
\\
&&
T_2 = -\frac{t}{\mu x},
\\
&&
X=
-\frac{1}{2\mu}
\left( 
\frac{1}{\mu x} -\mu x
\right)
-\frac{t^2}{2x},
\end{eqnarray}
into eqn(\ref{33}) yields
\begin{eqnarray}
ds^2
=
\frac{1}{\mu^2 x^2} (-dt^2 +dx^2 ).\label{2}
\end{eqnarray}
In this chart the coordinate variables take values
 between $-\infty<x<0,\ -\infty<t<\infty $ and the spatial
infinity stays at $x=-0$.

Next we  precisely define asymptotic anti-de sitter (AAdS) space  
in two dimensions in the same spirit of 
the three dimensional one proposed by Brown and Henneaux\cite{BH}.

To fix boundary conditions of AAdS spaces,
 we must analyze a little bit what should be a AAdS space.
 Let us divide a metric of  a will-be AAdS space  
 into the AdS background metric (\ref{2})
and its deviation near the spatial infinity ($x \sim -0$).
\begin{eqnarray}
&&
g_{ab} =\bar{g}_{ab}+h_{ab}, 
\\
&&
\bar{g}_{ab}dx^a dx^b =\frac{1}{\mu^2 x^2} (-dt^2 +dx^2) .
\end{eqnarray}
The asymptotic form of $h_{ab}$ near $x=-0$ 
 should be specified by some boundary conditions  
 for the space to be AAdS. 
Also it must be stressed that the boundary conditions 
 are invariant under the $SO(1,2)$ AdS isometry. This is
 a crucial point clarified by Brown and Henneaux. 
Thus let us AdS transform the metric at the infinity by using 
eqns(\ref{3})(\ref{34})(\ref{35}).
\begin{eqnarray}
\delta_\epsilon h_{ab} &=&
\nabla_a \epsilon_b +\nabla_b \epsilon_b
\nonumber\\
&=&
\bar{g}_{ac}\bar{\nabla}_b \epsilon^c+
\bar{g}_{bc}\bar{\nabla}_a \epsilon^c
\nonumber\\
&&
+
\epsilon^c \bar{\nabla}_c h_{ab}
+
h_{ac}\bar{\nabla}_b \epsilon^c+
h_{bc}\bar{\nabla}_a \epsilon^c
\nonumber\\
&&
+o(h^2 ).
\end{eqnarray}
This realtions take simple forms in the conformal coodinates:
\begin{eqnarray}
\delta_\epsilon h_{\pm \pm}
&=&
2h_{\pm\pm} \partial_\pm \epsilon^\pm
+\epsilon^+ \partial_+ h_{\pm\pm}
+\epsilon^- \partial_- h_{\pm\pm},
\label{36}\\
\delta_\epsilon h_{+-}
&=&
\left(-\frac{1}{2\mu^2 x^2}+h_{+-} \right)
\left[
\partial_+ \epsilon^+ +\partial_- \epsilon^-
-\frac{\epsilon^+ -\epsilon^-}{x}
\right]
\nonumber\\
&&
+\epsilon^+  
\partial_+ h_{+-} 
+\epsilon^-
\partial_- h_{+-} 
+\frac{\epsilon^+ -\epsilon^-}{x}h_{+-}.
\label{5}
\end{eqnarray}

It is important to note in eqn (\ref{5})  that
\begin{eqnarray}
&&
\lim_{x\rightarrow 0}\left( -\frac{1}{2\mu^2 x^2}
\left[
\partial_+ \epsilon^+ +\partial_- \epsilon^-
-\frac{\epsilon^+ -\epsilon^-}{x}
\right]
\right)
\nonumber\\
&&
=-\frac{1}{3\mu^2 }\frac{d^3 }{dt^3}
\left[
\epsilon_0  +\epsilon_1 \sin(\mu t+\delta_\epsilon )
\right] =o(1).
\end{eqnarray}
This quantity is of the zeroth order of $x$ near $x=-0$ 
and thus nonvanishing in the asymptotic region. 
Consequently the weakest boundary condition for 
the $g_{+-}$ component of AAdS metric should take a form
 as
\begin{eqnarray}
g_{+-}=-\frac{1}{2\mu^2 x^2} +o(1 ). \label{7}
\end{eqnarray}

Inspired by the Brown-Henneaux argument, 
we next impose another constraint 
on the AAdS boundary conditon search. 
 We will discuss later the AdS gravity model in which spacetimes
 of the solutions is the AdS space with $R=-2\mu^2$. 
Thus the model always 
possesses the two dimensional BTZ blackhole solution: 
\begin{eqnarray}
ds^2 =-\left(\mu^2 r^2 -a \right) dt^2
+\frac{dr^2}{\mu^2 r^2 -a} 
\end{eqnarray}
where $a$ is a constant related with mass. 
To map $r=\infty$ to $x=-0$, we change the coordinate as
$$
x=-\frac{1}{\mu^2 r}.
$$
Then the metric behaves near $x=-0$ like
\begin{eqnarray}
ds^2 &=& \frac{1}{\mu^2 x^2}(-dt^2 +dx^2 )
+a(dt^2 +dx^2) +o(x^2)
\nonumber\\
&=&
-\frac{dx^+ dx^-}{\mu^2 x^2}
+\frac{a}{2}(dx^{+2} +dx^{-2} )
+o(x^2).
\label{6}
\end{eqnarray}
The second constraint is that this solution(\ref{6}) must also 
 be AAdS.  This is a just similar requirement to the three 
dimensional one \cite{BH} that point particle
  solutions should be AAdS. 
Taking account of eqns(\ref{36})(\ref{6}), the AAdS
 boundary condition for the $g_{\pm\pm}$ component is specified
 as
\begin{eqnarray}
g_{\pm\pm}=o(1).\label{8}
\end{eqnarray}

Next we argue a possibility of asymptotic isometry extension.
 When some of general coordinate transformations 
make the AAdS metric boundary conditions(\ref{7})(\ref{8}) 
 invariant, the symmetry is called 
asymptotic isometry of the AdS space. 
From the two dimensional nature one can notice a chance  
 for the asymptotic isometry  to be raised to a Virasoro symmetry. 
Let us consider a conformal transformation:
\begin{eqnarray}
x^{'\pm}=F(x^\pm ),\label{500}
\end{eqnarray}
where $F(x)$ is an arbitrary function.
Then it can be shown that 
metrices satisfing the boundary condition (\ref{7})(\ref{8})
 are transformed into those which also satisfy the same conditions 
(\ref{7})(\ref{8}):
\begin{eqnarray}
ds^2
&=&
-\frac{dx^{'+} dx^{'-}}{\mu^2 x^{'2}}
+o(1) dx^{'+ 2} +o(1)dx^{'-2} +o(1) dx^{'+} dx^{'-}
\nonumber\\
&=&
-dx^+ dx^-
\left( 
\frac{1}{\mu^2 x^2}
+
\frac{1}{\mu^2}
\left[
\frac{2}{3}\frac{\frac{d^3 F(t)}{dt^3}}{\frac{dF(t)}{dt}}
-\left( 
\frac{\frac{d^2 F(t)}{dt^2}}{\frac{dF(t)}{dt}}
\right)^2
\right]\right)
\nonumber\\
&&
+o(1) dx^{+ 2} +o(1) dx^{-2} +o(1) dx^{+} dx^{-}
+o(x^2).
\nonumber
\end{eqnarray}
Thus from the observation of asymptoticity of AAdS metrices, 
the asymptotic isometry can be conjectured 
 the Virasoro symmetry (\ref{500}) 
to which  the $SO(1,2)$ AdS isometry (\ref{501}) 
belongs as a subgroup.

This conjecture sounds quite fascinating. 
But  rigorously speaking,
 in order to check whether the asymptotic conformal symmetry 
really survives or not, detailed analysis 
of the dynamics is needed. 
In fact, as we will argue later, the Virasoro symmetry 
 does not realize in the following AdS dilaton gravity. Moreover
 even for the $SO(1,2)$ AdS isometry it is impossible to
construct a nonsinglet representation 
except JT model.

\section{AdS Dilaton Gravity}
\ \\

Because 
Einstein gravity in two dimensions has trivial dynamics,
some matter fields are requested 
if one wants to construct meaningful gravitational models. 
Here we introduce a scalar dilaton 
field  and write down the most general action 
of the models in which 
all spacetime solutions are the AdS space with $R=-2\mu^2$. 
The models are called AdS dilaton gravity. 
Imposing renomalizability, the action reads
\begin{eqnarray}
S=
\frac{1}{16\pi G}
\int d^2 x \sqrt{-g}
\left[
\phi R
+
4\lambda (\nabla \phi)^2
+\frac{\mu^2}{4\lambda}
\left(
U_o e^{8\lambda \phi} -1
\right)
\right],
\end{eqnarray}
where $\phi$ is dilaton field, $G$  the gravitational constant,
 $\lambda$ and $U_o$ are real constants. 
When $U_o =0$ and negative $\lambda$,  
the model is reduced into the Liouville
induced gravity model, as is well known. While if we take 
$U_o =1$ and $\lambda\rightarrow 0$, the model is equivalent
with JT model.

Equations of motion are easily derived from the action.
\begin{eqnarray}
R=
8\lambda \nabla^2 \phi -2\mu^2 U_o e^{8\lambda \phi},
\label{502}
\end{eqnarray}

\begin{eqnarray}
&&
\left( 
g_{ab} \nabla^2 -\nabla_a \nabla_b
\right)\phi
+4\lambda
\left[
\nabla_a \phi \nabla_b \phi 
-\frac{1}{2}g_{ab} (\nabla \phi)^2
\right]
\nonumber\\
&&
=\frac{\mu^2}{8\lambda} 
\left[
U_o e^{8\lambda \phi} -1
\right]g_{ab} .
\label{503}
\end{eqnarray}
By taking the trace of eqn(\ref{503}),  
 we obtain 
$$
\nabla^2 \phi=\frac{\mu^2}{4\lambda}
\left[
U_o e^{8\lambda \phi} -1
\right].
$$
Substitution of this into eqn (\ref{502}) clarifies 
 the nature of the AdS gravity:
\begin{eqnarray}
R=-2\mu^2 .
\end{eqnarray}

In order to find solutions,  
let us adopt the conformal gauge for convenience.
\begin{eqnarray}
&&
x^\pm =t\pm x ,
\\
&&
ds^2 =-e^{2\rho (x^+ ,x^- )} dx^+ dx^- .
\end{eqnarray}
Then various useful relations are available as follows. 
\begin{eqnarray}
&&
\Gamma^+_{++} =2\partial_+ \rho,
\\
&&
\Gamma^-_{--}=2\partial_- \rho,
\\
&&
\nabla^2 = -4e^{-2\rho} \partial_+ \partial_-,
\\
&&
R=8e^{-2\rho} \partial_+ \partial_- \rho .
\end{eqnarray}

In the conformal gauge the equations of motion are reexpressed as
\begin{eqnarray}
&&
\partial_+ \partial_- \rho +\frac{\mu^2}{4} e^{2\rho} =0,
\\
&&
\partial_+ \partial_- \phi
+\frac{\mu^2}{16\lambda}
e^{2\rho}
\left[
U_o e^{8\lambda \phi} -1
\right]
=0,
\\
&&
\partial_\pm^2 \phi
-2\partial_\pm \rho \partial_\pm \phi
-4\lambda (\partial_\pm \phi)^2
=0.
\end{eqnarray}

Firstly we concentrate on the case with 
 $U_o\neq 0$ and $\lambda \neq 0$. Then 
general static solutions in the conformal gauge 
are found with ease as follows.
\begin{eqnarray}
&&
e^{2\rho} =\frac{M^2}{\mu^2 \sin^2 (M x) } ,
\\
&&
\phi
=-\frac{1}{4\lambda}
\ln \left[
A \cot (Mx) +B
\right],
\end{eqnarray}
where $B$ is a real constant, $A$ and $M$ are 
real or pure imaginary constants satisfing 
$A^2 +B^2 =U_o $. Note that when $U_o <0$ one cannot get 
 any metric form in a full covered chart 
 with a real and static dilaton configuration.
Because we have interest only in models  with  static 
 background solutions, 
only positive $U_o$ case is later concentrated on.
 In this case  
a simple background solution is written explicitly as follows.
\begin{eqnarray}
e^{2\rho} &=&\frac{1}{ \sin^2 (\mu x) },
\\
\phi
&=&-\frac{1}{4\lambda}
\ln \left[
-\sqrt{U_o} \cot (\mu x) 
\right]
\nonumber\\
&=&
-\frac{1}{4\lambda} \ln \left[\frac{\sqrt{U_o}}{-\mu x} \right]
+o(x^2).
\end{eqnarray}

We can also obtain general solutions in the conformal gauge:
\begin{eqnarray}
&&
e^{2\rho}
=
\frac{\dot{g}_+ (x^+ ) \dot{g}_- (x^- )}{ \left[ 
1+\frac{\mu^2 }{4}g_+ (x^+ )g_- ( x^-)
\right]^2},
\label{60}\\
&&
\phi
=
\frac{1}{8\lambda}\ln \left(\frac{4ab}{\mu^2 U_o} \right)
-\frac{1}{4\lambda}
\ln
\left[
\frac{a g_+ (x^+ ) - b g_- (x^-) +c}
{1+\frac{\mu^2}{4}g_{+}( x^+)g_- ( x^- )}
\right].\label{40}
\end{eqnarray}
The most general solutions are obtained by 
  coordinate transforming solutions (\ref{60})(\ref{40}) arbitrarily.

Here it should be commented that 
 because the candidate of AAdS isometry (\ref{500}) 
is a kind of conformal symmetry, 
 the solutions in the conformal gauge 
form a partial representation sector of the 
candidate group (\ref{500})  and 
can be investigated separately from  nonconformal 
solutions from the group theoretical point of view. 
 Moreover it can be shown straightforwardly 
 that the results obtained in the sector 
remain true even for other nonconformal solutions. 
Therefore we focus later 
on what happens in the conformal gauge sector
 of the configuration space.

Among solutions of the sector, 
 general AAdS solutions are picked up as follows.
\begin{eqnarray}
&&
e^{2\rho}
=
\frac{4}{\mu^2}
\frac
{
\dot{F} (x^+) \dot{F} (x^- )
}
{
\left[
e^{\frac{1}{2}(F(x^+) -F(x- ) )}
-e^{-\frac{1}{2}(F(x^+) -F(x- )) }
\right]^2
},
\\
&&
\phi
=
\frac{1}{8\lambda}\ln \left(\frac{AB}{ U_o} \right)
-\frac{1}{4\lambda}
\ln
\left[
\frac{
A e^{ F(x^+)} +B e^{-F(x^-)} 
 +C}
{1-e^{F(x^+) -F(x^- ) }}
\right].
\label{44}
\end{eqnarray}
It is easy to see that the asymptotic forms:
\begin{eqnarray}
&&
e^{2\rho}
=
\frac{1}{\mu^2 x^2}
\left[
1+x^2 
\left( 
\frac{2}{3}\frac{ \frac{d^3 F}{dt^3} }{ \frac{dF}{dt} }
-\left(\frac{ \frac{d^2 F}{dt^2} }{ \frac{dF}{dt}} \right)^2
-\frac{1}{3}\left(\frac{dF}{dt} \right)^2
\right)
+o(x^4)
\right],
\nonumber
\\
&&
\phi 
=
-\frac{1}{4\lambda}
\ln
\left[
\sqrt{\frac{U_o}{AB}}
\frac{Ae^{F(t)}+Be^{-F(t)} +C}{2(-x)\dot{F}}
\right]
+\frac{1}{4\lambda}
\frac{C\dot{F}}{Ae^{F(t)} +Be^{-F(t)} +C} x
+o(x^2 ).\label{21}
\nonumber
\end{eqnarray}
 really satisfy the AAdS conditions (\ref{7})(\ref{8}).

Next we make up canonical formulation of the model.
Firstly let us adopt ADM decomposition of the metric.
$$
ds^2 
=
-N^2 dt^2
+a^2 (dx +Jdt )^2 .
$$

\begin{eqnarray}
g_{ab}
=\left[
\begin{array}{cc}
-N^2 +a^2 J^2 & a^2 J \\
a^2 J & a^2
\end{array}
\right].
\end{eqnarray}

Then inverse of the metric is given as 
\begin{eqnarray}
g^{ab}
=\left[
\begin{array}{cc}
-\frac{1}{N^2} & \frac{J}{N^2} \\
\frac{J}{N^2} & \frac{1}{a^2} -\frac{J^2}{N^2}
\end{array}
\right].
\end{eqnarray}

Also extrinsic curvature is defined as follows.
\begin{eqnarray}
&&
K=\frac{1}{N}\left[ 
\partial_x J +\frac{J}{a}\partial_x a -\frac{\dot{a}}{a}
\right].
\end{eqnarray}

Then the action is reduced into the following form.
\begin{eqnarray}
S&=&
\frac{1}{8\pi G}\int d^2 x {\it L},
\nonumber\\
{\it L}
&=&
-\frac{1}{N}\left[
\dot{a} -\partial_x (aJ)
\right]
\left[
\dot{\phi} -J \partial_x \phi
\right]
+\frac{1}{a} \partial_x N \partial_x \phi
\nonumber\\
&&
-2\lambda \frac{a}{N}\left( 
\dot{\phi} -J\partial_x \phi
\right)^2
+2\lambda \frac{N}{a} \left(\partial_x \phi \right)^2
\nonumber\\
&&
+\frac{\mu^2 Na}{8\lambda}
\left( 
U_o e^{8\lambda \phi} -1
\right)
+surface\ term. \label{11}
\end{eqnarray}

From eqn(\ref{11}) conjugate momentums are defined in the usual way.
\begin{eqnarray}
&&
\Pi_a = -\frac{1}{8\pi GN}
\left[
\dot{\phi} -J \partial_x \phi
\right],
\nonumber\\
&&
\Pi_\phi
=
-\frac{1}{8\pi GN}
\left[
\dot{a} -\partial_x (aJ)
+4\lambda a
\left(
\dot{\phi} -J \partial_x \phi
\right)
\right].
\nonumber
\end{eqnarray}
Hamiltonian and momentum constraints arise 
from differentiation of the action 
with respect to lapse and shift functions.
\begin{eqnarray}
{\cal H}_N
&=&-\frac{\delta S}{\delta N}
\nonumber\\
&=&
-8\pi G \Pi_a \left( \Pi_\phi -2\lambda a \Pi_a \right)
\nonumber\\
&&
+\frac{1}{8\pi G}
\left[
\partial_x \left( 
\frac{1}{a}\partial_x \phi
\right)
-\frac{2\lambda}{a} (\partial_x \phi)^2
-\frac{\mu^2 a}{8\lambda}
\left( 
U_o e^{8\lambda \phi} -1
\right)
\right] \approx 0,
\nonumber\\
{\cal H}_J
&=&-\frac{\delta S}{\delta J}
\nonumber\\
&=&
\Pi_\phi \partial_x \phi
-a \partial_x \Pi_a
\approx 0.
\nonumber
\end{eqnarray}

It is also known that 
the  canonical equations of motion come from
the action:
\begin{eqnarray}
S_{can}
=\int dt \left( 
\int^0_{-\infty} dx 
\left(\Pi_a \dot{a} +\Pi_\phi \dot{\phi}
\right)
-H[\epsilon^t =1,\epsilon^x =0] 
\right),
\end{eqnarray}
where 
\begin{eqnarray}
&&
H[\epsilon]
=
\int^0_{-\infty} dx 
\left( 
\epsilon^N {\cal H}_N +\epsilon^J {\cal H}_J
\right)
+Q[\epsilon],
\\
&&
\epsilon^N =N\epsilon^0 ,
\\
&&
\epsilon^J
=\epsilon^1 +J\epsilon^0 ,
\end{eqnarray}
\begin{eqnarray}
Q[\epsilon ]
&=&
\left[\frac{1}{8\pi G \bar{a}^2}
\epsilon^{\bar{N}}  
\partial_x \bar{\phi} (a-\bar{a}) 
\right]_{x=-0}
\nonumber\\
&&
+\left[\frac{1}{8\pi G \bar{a}}
\left[\epsilon^{\bar{N}}\left(
-\partial_x ( \phi-\bar{\phi}) 
+4\lambda  \partial_x \bar{\phi} (\phi-\bar{\phi})
\right)
+\partial_x \epsilon^{\bar{N}}  (\phi-\bar{\phi}) 
\right]\right]_{x=-0} 
\nonumber\\
&&
+\left[\epsilon^{\bar{J}}
\left[
\bar{a}( \Pi_a -\bar{\Pi}_a )
-\bar{\Pi}_\phi (\phi -\bar{\phi}) 
\right]\right]_{x=-0},
\nonumber
\\
&&
\epsilon^{\bar{N}} =\bar{N}\epsilon^0 ,
\nonumber\\
&&
\epsilon^{\bar{J}}
=\epsilon^1 +\bar{J}\epsilon^0 .\nonumber
\end{eqnarray}
Here we need boundary conditions how the dynamical variables
 approach to the background values.
For the general solutions of the dilaton field, we make  
 dilaton configurations restricted as
\begin{eqnarray}
\phi
=
-
\frac{1}{4\lambda} \ln \left(\frac{\sqrt{U_o}}{-\mu x} \right)
+
\sum_{n=0}\phi_n (t) x^n .\label{20} 
\end{eqnarray}
Unfortunately it is not known apriori 
how fast the dilaton field approches to the background value. 
However, there are two typical cases as follows.
Firstly we set in general  $\phi_0 \neq 0$ in eqn (\ref{20}) . 
Taking account of AAdS conditions of the metric, 
the boundary condition
is expressed as follows.
\begin{eqnarray}
&&
N= -\frac{1}{\mu x} +o(x) ,
\\
&&
a=-\frac{1}{\mu x} +o(x),
\\
&&
J=o(x^2),
\\
&&
\phi = \frac{1}{4\lambda} \ln \left(
\frac{\sqrt{U_o}}{-\mu x}
\right) 
+\phi_0 (t) + o(x).
\end{eqnarray}
This boundary condition is characterized by a fact that  
action of the asymptotic isometry candidate (\ref{500}) on 
the specified configuration space is close, that is,
\begin{eqnarray}
\phi &=&
\frac{1}{4\lambda}
\ln
\left[
-\frac{2\sqrt{U_o}}{\mu}
\frac{1}{F(x^+) -F(x^-)}
\right]
+\phi_o
\left(\frac{1}{2}(F(x^+ )+F(x^-) \right) +o(x)
\nonumber\\
&=&
\frac{1}{4\lambda}\ln
\left(\frac{\sqrt{U_o}}{-\mu x} \right)
+\phi_o (F(t)) -\frac{1}{4\lambda}\ln \dot{F} (t) +o(x).
\end{eqnarray}

From the boundary condition, 
 variations of the dynamical variables in order 
to get the equations
 of motion are also constrained.
\begin{eqnarray}
&&
\delta a = o(x),
\\
&&
\delta \phi =\delta\phi_0 (t) +o(x),
\\
&&
\delta \Pi_a =o(x),
\\
&&
\delta \Pi_\phi =o(1).
\end{eqnarray}
By making the action stationary, we get just one boundary 
 equation of motion from the action:
\begin{eqnarray}
\phi_1 (t)=0.
\end{eqnarray}
This condition means that
$$
C=0
$$
 in eqn(\ref{44}).

Next let us discuss the second boundary 
condition in which $\phi_0$ is fixed to $0$,
 but $\phi_1$ does not vanish in general. 
\begin{eqnarray}
&&
N= -\frac{1}{\mu x} +o(x),
\\
&&
a=-\frac{1}{\mu x} +o(x),
\\
&&
J=o(x^2),
\\
&&
\phi = 
-\frac{1}{4\lambda} 
\ln \left( 
\frac{\sqrt{U_o}}{-\mu x} 
\right)
 +\phi_1 (t) x +o(x^2).
\end{eqnarray}
It is noticed that 
because  both AdS and Virasoro transformation 
(\ref{501})(\ref{500}) 
 generate a nonvanishing $\phi_0$ term,
  the condition breaks the symmetries explicitly. 
The variational conditions read
\begin{eqnarray}
&&
\delta a = o(x),
\\
&&
\delta \phi =\delta\phi_1 (t) x +o(x^2),
\\
&&
\delta \Pi_a =o(x^2),
\\
&&
\delta \Pi_\phi =o(x).
\end{eqnarray}
In this case we do not have any nontrivial equation of motion 
on the boundary from the action.

For the above two cases 
 let us consider charges of the asymptotic isometry candidate
 (\ref{500}).
\begin{eqnarray}
Q[\epsilon] &=&
\frac{1}{16\pi G} 
\left[
e^{-2\bar{\rho}(x)}\dot{\bar{\phi}}\left[ 
\epsilon^t ( g_{xx} -\bar{g}_{xx})
 +2\epsilon^x ( g_{tx} -\bar{g}_{tx})
\right]
\right]_{x=-0}
\nonumber\\
&&
-\frac{1}{8\pi G}
\left[
\left( 
\epsilon^t \partial_x +\epsilon^x \partial_t
-\epsilon^t (\dot{\bar{\rho}} +4\lambda \dot{\bar{\phi}} )
-(\partial_x \epsilon^t)
\right)
( \phi -\bar{\phi})
\right]_{x=-0},
\end{eqnarray}
where
\begin{eqnarray}
&&
e^{2\bar{\rho}} =\frac{1}{\sin^2 (\mu x)} ,
\\
&&
\bar{g}_{ab}dx^a dx^b =e^{2\bar{\rho}}(-dt^2 +dx^2 ),
\\
&&
\bar{\phi} =-\frac{1}{4\lambda}
\ln
\left( 
\frac{\sqrt{U_o}}{-\mu x}
\right),
\\
&&
\epsilon^t =\frac{1}{2}[\epsilon (x^+) +\epsilon (x^- )],
\\
&&
\epsilon^x = \frac{1}{2}[\epsilon (x^+ )-\epsilon (x^- )].
\end{eqnarray}
Substituting the AAdS solutions into $Q$ and manipulating
 a littel bit, we acquire for the both cases
\begin{eqnarray}
Q[\epsilon]&=& -\frac{\epsilon (t)}{8\pi G} 
\partial_x \delta \phi (x=-0,t)
\nonumber\\
&=&
-\frac{\epsilon (t)}{32\pi G\lambda}
\frac{ C\dot{F}}
{Ae^F +Be^{-F} +C}.
\end{eqnarray}

For the first boundary condition: $\phi_0 \neq 0$,
all the charges of the Virasoro transformation (\ref{500}) 
vanish exactly: $Q[\epsilon]=0$ due to $C=0$. 
Thus the Virasoro symmetry (\ref{500}) realizes as AAdS isometry,
 however, the representation is just singlet and trivial.

In the second case with $\phi_0 =0$, time independence 
of the charges demands that
\begin{eqnarray}
&&
C\neq 0,
\\
&&
\epsilon (t)=const,
\\
&&
\frac{ \dot{F}}
{Ae^F +Be^{-F} +C} =\frac{\mu}{2\sqrt{AB}}.
\end{eqnarray}
Thus, as expected, it is concluded that 
the asymptotic isometry includes  only time translation. 
 Then the form of AAdS solution is
 constrained as
\begin{eqnarray}
&&
e^{F(t)}
=
-\frac{1}{2A}
\left[
C+\sqrt{C^2 -4AB}\tanh
\left( 
\frac{\mu}{4\sqrt{AB}}\sqrt{C^2 -4AB} (t+t_o)
\right)
\right],
\nonumber\\
&&
\phi
=-\frac{1}{4\lambda}
\ln \left(
\frac{\sqrt{U_o}}{-\mu x} 
\right)
+\frac{\mu}{16\lambda}
\frac{C}{\sqrt{AB}}x +o(x^2) \nonumber,
\end{eqnarray}
and the conserved energy is given as follows.
\begin{eqnarray}
E=- \frac{\mu }{64\pi G\lambda}\frac{C}{\sqrt{AB}}.
\end{eqnarray}

For negative $\lambda$ and $U_o =0$, the model
 is reduced into the Liouville theory with negative 
cosmological constant. 
It is a well known fact that 
by integrating out the dilaton field in the action:
\begin{eqnarray}
S=
\frac{1}{16\pi G}
\int d^2 x \sqrt{-g}
\left[
\phi R
-4|\lambda|(\nabla \phi)^2
+\frac{\mu^2}{4|\lambda|}
\right]
\end{eqnarray}
 the original 
 polyakov action\cite{P} appears:
\begin{eqnarray}
S_p =
\frac{1}{16\pi G}
\int
d^2 x \sqrt{-g}
\left[
-\frac{1}{16|\lambda|} R\frac{1}{\nabla^2}R
+\frac{\mu^2}{4|\lambda|}
\right].
\end{eqnarray}

In the Liouville case, any static background solution
with 
$$
e^{2\rho} =\frac{1}{\sin^2 (\mu x)}
$$
 cannot exist and the dilaton field has explicit time 
dependence. However this difference 
between the cases with $U_o \neq 0$ 
and the Liouville model is not so significant, 
and  does not change the results 
for the case with $U_o \neq 0$ 
 if we take another background solution: 
\begin{eqnarray}
&&
d\bar{s}^2 = -\frac{1}{\mu^2 x^2} dx^+ dx^-,
\\
&&
\bar{\phi}= -4\ln (-\mu x).
\end{eqnarray}
Repeating the same analysis of the $U_o\neq 0$ cases, 
it is easily shown that 
the Liouville theory with negative cosmological constant
  have a singlet representaion of Virasoro 
algebra (\ref{500}) as  AAdS isometry and a representation 
 where only time translation is preserved.

\section{JT Model}

When taking $U_o =1$ and $\lambda =0$ the model expresses
the JT model:
\begin{eqnarray}
S=
\frac{1}{16\pi G}
\int d^2 x \sqrt{-g}
\phi (R+2\mu^2 ).
\end{eqnarray}

In the case general AAdS solutions in the conformal gauge
 are also obtained.
\begin{eqnarray}
e^{2\rho}
&=&
\frac{4}{\mu^2}\frac{\dot{F}(x^+) \dot{F}(x^-)}
{\left[ 
e^{
\frac{1}{2}
\left( 
F(x^+) -F(x^- )
\right)
}
-
e^{-
\frac{1}{2}
\left( 
F(x^+) -F(x^- )
\right)
}
\right]^2 },
\label{50}\\
\phi
&=&
A\frac{1+e^{F(x^+) -F(x^- )}}
{1-e^{F(x^+) -F(x^- )}}
\nonumber\\
&&
+
B\frac{1}
{e^{-F(x^+) }-e^{ -F(x^- )}}
+
C
\frac{1}
{e^{F(x^- )}-e^{F(x^+) }}.
\label{51}
\end{eqnarray}
They behave near $x=-0$ like  
\begin{eqnarray}
e^{2\rho}
&=&
\frac{1}{\mu^2 x^2}+
\frac{1}{\mu^2}
\left( 
\frac{2}{3}\frac{1}{\dot{F}}\frac{d^3 F}{dt^3}
-\frac{\ddot{F}^2}{\dot{F}^2}
-\frac{1}{3}\dot{F}^2
\right)
+o(x^2),
\nonumber\\
\phi
&=&
-\frac{1}{2\dot{F} x}
\left(2A+Be^F +Ce^{-F} \right)
\nonumber\\
&&
+\frac{x}{12\dot{F}}
\left[
\frac{1}{\dot{F}}\frac{d^3 F}{dt^3}
\left(2A+Be^F +Ce^{-F} \right)
\right.\nonumber\\
&&\left.\ \ \ \ \ \ \ \ \ \ \ \ 
-3\ddot{F}\left(Be^F -Ce^{-F} \right)
-\dot{F}^2 \left(4A-Be^F -Ce^{-F} \right)
\right]
+o(x^3).
\nonumber
\end{eqnarray}
In later analysis we take the following
 static solution among them as a background 
which AAdS solutions are approaching to.
\begin{eqnarray}
d\bar{s}^2 &=&\bar{g}_{ab}dx^a dx^b=e^{2\bar{\rho}}(-dt^2 +dx^2 )
\nonumber\\
&=& \frac{1}{\sin^2 (\mu x)}\left(-dt^2 +dx^2 \right),
\label{22} \\
\bar{\phi}&=&-\cot (\mu x) =-\frac{1}{\mu x} +o(x).
\label{23}
\end{eqnarray}
The charge of the asymptotic isometry candidate (\ref{500})
 is expressed in this case as
\begin{eqnarray}
Q[\epsilon] &=&
\frac{1}{16\pi G} 
\left[
e^{-2\bar{\rho}(x)}\dot{\bar{\phi}}\left[ 
\epsilon^t ( g_{xx} -\bar{g}_{xx})
 +2\epsilon^x ( g_{tx} -\bar{g}_{tx})
\right]
\right]_{x=-0}
\nonumber\\
&&
-\frac{1}{8\pi G}
\left[
\left( 
\epsilon^t \partial_x +\epsilon^x \partial_t
-(\partial_x \epsilon^t)
\right)
( \phi -\bar{\phi})
\right]_{x=-0},
\end{eqnarray}
where
\begin{eqnarray}
&&
\epsilon^t =\frac{1}{2}[\epsilon (x^+) +\epsilon (x^- )],
\nonumber\\
&&
\epsilon^x = \frac{1}{2}[\epsilon (x^+ )-\epsilon (x^- )].
\nonumber
\end{eqnarray}

Just as $U_o \neq 0$ cases,
 we consider two independent boundary condition of the
 dilaton field. Near $x=-0$ the canonical variables 
 are expanded with respect to $x$ as follows.
\begin{eqnarray}
&&
a=-\frac{1}{\mu x} +a_1 (t)x +o(x^2 )
\\
&&
N=-\frac{1}{\mu x} +N_1 (t) x +o(x^2)
\\
&&
J=J_2 (t) x^2 +o(x^3)
\\
&&
\phi = -\frac{\phi_{-1} (t)}{\mu x} +x \phi_1 (t)
+o(x^2)
\label{70}
\\
&&
\Pi_a =\frac{1}{8\pi G}
\left[
-\dot{\phi}_{-1}
+(\mu \dot{\phi}_0 -J_2 \phi_{-1} )x
+o(x^2)
\right]
\\
&&
\Pi_\phi =
\frac{1}{8\pi G}\left[ 
J_2 x +\mu \dot{a}_1 x^2 +o(x^3)
\right]
\end{eqnarray}
Here a zeroth order term $\phi_0(t)$ in the dilaton field 
is omitted in order to make the charge $Q$ finite. 
For the first example, 
$\phi_{-1}$ is not fixed to any specified value in the 
configration space. This condition is naturally required 
if one repects the AAdS isometry candidate (\ref{500})
 in this level because the dilaton
 solution (\ref{23}) is  transformed as
\begin{eqnarray}
&&
\delta_\epsilon \phi=\frac{\dot{\epsilon}(t)}{\mu x}+o(x)
=\frac{o(1)}{x}+o(x) ,
\\
&&
x^{\pm'}=x^\pm - \epsilon (x^\pm ).
\end{eqnarray}
Adopting the boundary conditions, 
the boundary equations of motion demand that
\begin{eqnarray}
&&
a_1 =\bar{a}_1 =-\frac{\mu}{6},
\\
&&
N_1 =\bar{N}_1 =-\frac{\mu}{6},
\\
&&
\phi_{-1} =1.
\end{eqnarray}
These require that
\begin{eqnarray}
&&
\dot{F}=\frac{\mu}{2}
\left(2A+Be^F +Ce^{-F} \right),
\\
&&
BC-A^2 =1,
\end{eqnarray}
in eqns(\ref{50})(\ref{51}).
Consequently 
this yields 
the vanishing charges of the conformal transformation:
$$
Q[\epsilon]=0.
$$
Thus the Virasoro symmetry (\ref{500}) including the AdS 
isometry (\ref{501}) survives, but the representation remains
 trivially singlet.

The second proposal for the boundary condition  is 
to fix the asymptotic dilaton field as
$$
\phi_{-1} =1
$$ 
in the configuration space level. This breaks explicitly 
the Virasoro symmetry (\ref{500}) except time translation. 
While no additional constraint arises from the boundary equation.
 Solutions belonging to the specified configuration space  must 
satisfy 
\begin{eqnarray}
&&
\dot{F}=\frac{\mu}{2}
\left(2A+Be^F +Ce^{-F} \right),
\\
&&
\phi = -\frac{1}{\mu x}
+\frac{\mu x}{3}(BC-A^2) +o(x^2) ,
\\
&&
e^{2\rho}=\frac{1}{\mu^2 x^2}
+\frac{1}{3}(BC -A^2) +o(x^3).
\end{eqnarray}
From these relations one can calculate the charge straightforwardly
 as follows.
$$
Q[\epsilon]=\frac{\mu \epsilon(t)}{16\pi G}
(A^2 -BC+1).
$$
Thus the repesentation of AAdS isometry is nonsinglet, 
but the symmetry  consists of only time translation and 
 only energy:
$$
E=\frac{\mu}{16\pi G} (A^2 -BC+1)
$$
is conserved.

Though the standard canonical representations of
 JT model cannot keep nontrivially the $SO(1,2)$ 
AdS  and the Virasoro symmetry (\ref{501})(\ref{500}) 
as shown above, 
 it is quite notable that 
 the JT model has a noncanonical but 
nonsinglet representaion of the $SO(1,2)$ 
AdS isometry (\ref{501}) with finite conserved charges.

Let us consider  AdS transformed backgrounds 
genetated by acting $SO(1,2)$ on  the reference solution(\ref{23}). 
By selecting the canonical variables 
approaching to the transformed backgrounds
  different canonical representation sectors are generated. 
Moreover let us adopt the first boundary condition(\ref{70})
 in each sector. 
Combining all the canonical sectors,  
 a hybrid representation for the total system can be defined. 
 The hamiltonian which expresses the time evolution
of the total system is written as follows.
\begin{eqnarray}
&&
H_{tot} =\sum_L 
{\bf 0}\otimes \cdots {\bf 0}\otimes
H[\epsilon^t =1,\epsilon^x =0,L]
\otimes{\bf 0} \cdots \otimes{\bf 0} 
\end{eqnarray}
where 
\begin{eqnarray}
&&
H[\epsilon ,L]
=\int
dx
\left(\epsilon {\cal H}_N +\epsilon^J {\cal H}_J \right)
+\tilde{Q}[\epsilon ,L] ,
\nonumber\\
\tilde{Q}[\epsilon ,L]
&=&
\frac{1}{8\pi G \bar{a}^2_L}
\left[
\epsilon^{\bar{N}_L} 
\left( 
\partial_x \bar{\phi}_L (a-a_v ) 
-\bar{a}_L\partial_x ( \phi-\phi_v ) 
\right)
+\partial_x \epsilon^{\bar{N}_L} 
\bar{a}_L (\phi-\phi_v ) 
\right]_{x=-0} 
\nonumber\\
&&
+\epsilon^{\bar{J}_L}
\left[
\bar{a}_L ( \Pi_a -\Pi_{av} )
-\bar{\Pi}_{\phi L} (\phi -\phi_v ) 
\right]_{x=-0}.
\nonumber
\end{eqnarray}
and the transformed  background qunatities 
by a $SO(1,2)$ element $L$ are denoted 
$\bar{a}_L$, $\bar{\phi}_L$ and so on. 
Note that  due to 
the isometry all the metric quantities of backgrounds 
 remain unchanged under the transformation, that is, 
$\bar{a}_L =\bar{a}$, $\bar{J}_L =\bar{J}$ 
and $\bar{N}_L =\bar{N}$. 
Here the vacuum in the representation is chosen as
\begin{eqnarray}
&&
ds^2_v =\frac{1}{\sin^2 (\mu x)} 
\left(-dt^2 +dx^2 \right),
\\
&&
\phi_v =0 .
\end{eqnarray}
Due to the boundary equations of motion,
 the singlet representation is automatically selected  and  
 all the solutions  
take the same value of $\tilde{Q}$ in each sector. 
 Thus the value of $\tilde{Q}$ for a certian solution  
depends only on which background sector the solution belongs to.

The charges of the AdS isometry can be written more explicitly as
\begin{eqnarray}
\tilde{Q}[\epsilon ,L]=
-\frac{1}{8\pi G}
\left[
\epsilon^t 
\left(
\partial_x +\frac{1}{x} -\frac{1}{3}\mu^2 x
 \right)
+\epsilon^x \partial_t -(\partial_x \epsilon^t )
\right]\bar{\phi}_L .
\end{eqnarray}
For the reference background solution (\ref{22})(\ref{23}), 
the charge is calculated straight forwardly as 
\begin{eqnarray}
\tilde{Q}[\epsilon,L={\bf 1} ]
=-\frac{\mu}{8\pi G}\epsilon_0.
\end{eqnarray}
Thus the three charges of the AdS symmetry 
can be conservsed. Among the charges of the solution,
 only energy is nonvanishing and takes a  negative value of
\begin{eqnarray}
E=\frac{\partial \tilde{Q}}{\partial \epsilon_0}
=-\frac{\mu}{8\pi G}.
\end{eqnarray}

Meanwhile, 
modifing  the reference solution (\ref{23}) 
 by the infinitesimal AdS transformtion,
the dilaton field responds as follows.
\begin{eqnarray}
&&
\delta_\xi \phi
=\frac{\dot{\xi}}{\mu x}
+\frac{x}{6\mu}
\left( 
\frac{d^3 \xi}{dt^3} +2\mu^2 \dot{\xi}
\right),
\\
&&
\xi (t) =
\xi_0 +\xi_1 
\sin
\left( 
\mu t +\delta _\xi
\right).
\end{eqnarray}
This yields  deviation of the charge 
from the value of the background 
(\ref{22})(\ref{23}) like
\begin{eqnarray}
\delta_\xi \tilde{Q}
=
\frac{\mu^2}{8\pi G}
\epsilon_1 \xi_1
\sin(\delta_\xi -\delta_\epsilon ) .
\end{eqnarray}
It means that the reference solution (\ref{22})(\ref{23})
 belongs to a nonsinglet representation of the AdS isometry.
To see it in a finite transformation, let us take 
a AdS transformation $L_1$ :
\begin{eqnarray}
&&
T_1 =T_1^{'} \cosh\omega -X' \sinh\omega ,
\\
&&
T_2 =T_2^{'},
\\
&&
X=X' \cosh\omega -T_1^{'} \sinh\omega,
\end{eqnarray}
in eqns (\ref{100})(\ref{33}). 
Then the dilaton field (\ref{23}) is transformed into
\begin{eqnarray}
\bar{\phi}_{L_1}
=
-\cot(\mu x) \cosh\omega
+\frac{\cos(\mu t)}{\sin(\mu x)}
\sinh\omega.
\end{eqnarray}
and its charge is evaluated as
\begin{eqnarray}
\tilde{Q}[\epsilon, L_1 ]
=
-\epsilon_0 \frac{\mu}{8\pi G} \cosh\omega
-\epsilon_1\sin \delta_\epsilon
\frac{\mu}{8\pi G} \sinh\omega.
\end{eqnarray}
Thus it has been verified that 
the AdS charges are 
 transformed exactly as a vector of the AdS transformation.

Therefore even though the AdS symmetry cannot be realized
 in each canonical representation sector, 
one can construct a nonsinglet representation 
by combining the sectors and defining the AdS transformation 
 from one sector to another by use of
\begin{eqnarray}
&&
\epsilon^t 
=\epsilon_0 +\frac{\epsilon_1}{2}
(\sin (\mu x^+ +\delta_\epsilon )
 +\sin(\mu x^- +\delta_\epsilon ) ) ,
\\
&&
\epsilon^x
=
\frac{\epsilon_1}{2}
(\sin (\mu x^+ +\delta_\epsilon )
 -\sin(\mu x^- +\delta_\epsilon ) ) ,
\\
&&
\delta \phi
=\epsilon^a \nabla_a \phi ,
\\
&&
\delta g_{ab} =\nabla_a \epsilon_b +\nabla_b \epsilon_a .
\end{eqnarray}

Though this hybrid representation is not purely 
canonical, we want to stress that  
it is a well defined representation 
with finite and  nonvanishing AdS casimir operator:
\begin{eqnarray}
M^2 = \left(\frac{\mu}{8\pi G} \right)^2.
\end{eqnarray}
Therfore it can be argued that JT model have a nontrivial 
representation of the AdS isometry.

Unfortunately the energy takes only negative values 
in the representation,  positive energy 
excitations of matter fields cannot be incorpolated. 
If one wants positive energy to take part in the dynamics,
 another representation discussed above 
in which the AdS isometry is explicitly broken must be used.


\begin{thebibliography}{100}

\bibitem{BH}
J.D.Brown and M.Henneaux,
Commun.Math.Phys.104,207-226(1986)

\bibitem{M}
J.Maldacena,hep-th/9711200

\bibitem{S}
A.Strominger, J.High Energy Phy.,02:009,(1998)



\bibitem{W}
E.Witten,hep-th/9802150,hep-th/9803131.

\bibitem{P}
A.M.Polyakov,Phys.Lett.,B103,207,(1981).

\bibitem{JT}
R.Jackiw, in Quantum Theory of Gravity,ed.S.Christensen
(Adam Hilger, Bristol,1984)p.403;
C.Teitelboim,
in Quantum Theory of Gravity,ed.S.Christensen
(Adam Hilger, Bristol,1984)p.327.

\bibitem{H}
M.Henneaux,Phys.Rev.Lett,54,959,(1985).

\bibitem{T}
H.Terao,Nucl.Phys.,B395,623,(1993).


\end{thebibliography}
\end{document}